\documentclass[preprint,showpacs,amsmath,amssymb,prb,aps]{revtex4}
\usepackage{graphicx}% Include figure files
\usepackage{dcolumn}% Align table columns on decimal point
\usepackage{bm}% bold math

\begin{document}

%\preprint{03/26/03 Draft}

\title{Phase Diagram for Magnetic Reconnection in Heliophysical, Astrophysical and Laboratory Plasmas}

\author{Hantao Ji$^1$ and William Daughton$^2$}
\affiliation{$^1$Center for Magnetic Self-Organization, Princeton Plasma Physics Laboratory, Princeton University, Princeton, New Jersey 08543 \\
$^2$Los Alamos National Laboratory, Los Alamos, New Mexico 87545}

\date{\today}

\begin{abstract}

Recent progress in understanding the physics of magnetic reconnection is conveniently summarized in terms of a phase diagram which organizes the essential dynamics for a wide variety of applications in heliophysics, laboratory and astrophysics. The two key dimensionless parameters are the Lundquist number and the macrosopic system size in units of the ion sound gyroradius. 
In addition to the conventional single X-line collisional and collisionless
phases, multiple X-line reconnection phases arise due to the presence of the plasmoid instability either in collisional
and collisionless current sheets. In particular, there exists a unique phase 
termed \lq\lq multiple X-line hybrid phase" where a hierarchy of collisional islands or 
plasmoids is terminated by a collisionless current sheet, resulting in a rapid coupling between the macroscopic and kinetic scales and a mixture of collisional and collisionless dynamics.
The new phases involving multiple X-lines and collisionless physics may be important for the emerging
applications of magnetic reconnection to accelerate charged particles beyond their thermal speeds.
A large number of heliophysical and astrophysical plasmas are surveyed and grouped in the phase diagram: 
Earth's magnetosphere, solar plasmas (chromosphere, corona, wind and tachocline), galactic plasmas (molecular clouds, interstellar media, accretion disks and their coronae, Crab nebula, Sgr A*, gamma ray bursts, magnetars), extragalactic plasmas (Active Galactic Nuclei disks and their coronae, galaxy clusters, radio lobes, and extragalactic jets). 
Significance of laboratory experiments, including a next generation reconnection experiment, is also discussed.

\end{abstract}

\maketitle

\section{Collisional and collisionless reconnection}

It has been a long held view that magnetic reconnection is primarily characterized by plasma collisionality. 
This is evidenced by the common uses of the resistive magnetohydrodynamic (MHD) models, 
which is parameterized solely by the dimensionless Lundquist number,
\begin{equation}
S \equiv {\mu_0 L_{CS} V_A \over \eta},
\label{Lund}
\end{equation}
as a starting point of the discussion for magnetic reconnection.
In Eq.(\ref{Lund}), $L_{CS}$ is the half length of the reconnecting current sheet, and can be taken as $L_{CS}=\epsilon L$ 
where $L$ is the plasma size and $0 \leq \epsilon \leq 1/2$ (the choices of $\epsilon$ are discussed in Sec.VI.A.).
$V_A$ is the Alfv\'{e}n velocity based on the {\it reconnecting} magnetic field component
and $\eta$ is the plasma resistivity due to Coulomb collisions.
The well-known Sweet-Parker model~\cite{sweet58,parker57} predicts reconnection rates 
as an explicit function of $S$,
\begin{equation}
{V_R\over V_A} = {1 \over \sqrt{S}},
\label{SP}
\end{equation}
where $V_R$ is the reconnection inflow speed.
When collisions are sufficiently infrequent or $S$ is sufficiently large, 
physics beyond resistive MHD becomes crucial~\cite{birn01}, leading to a fast reconnection rate
nearly independent of $S$.
A large body of the work in the past decades, therefore, has focused on reconnection either in collisional or collisionless limit as summarized by recent reviews \cite{zweibel09,yamada10}. 

\begin{figure}
\begin{center}
\includegraphics[width=4.5in]{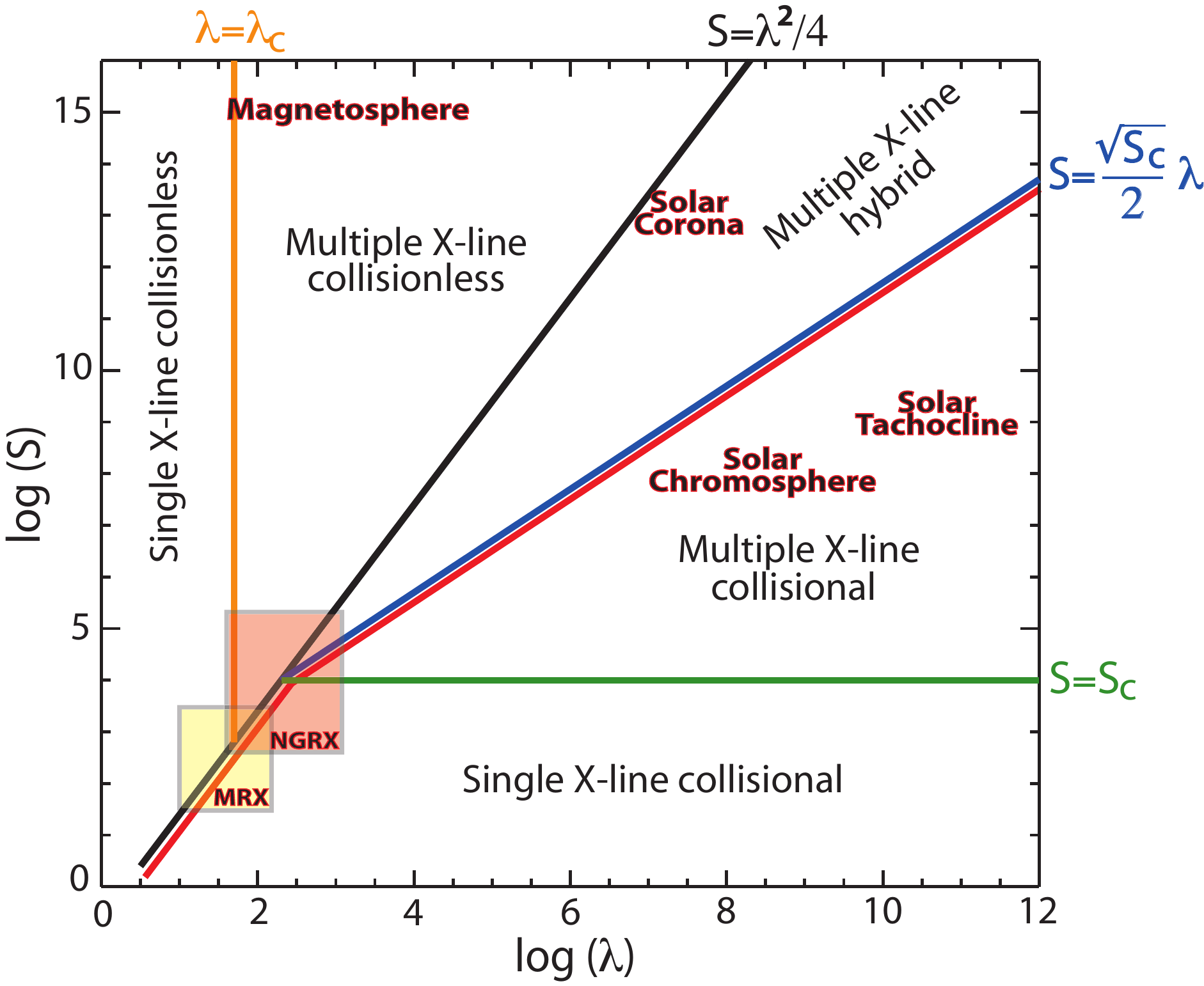}
 \caption{A phase diagram for magnetic reconnection in two dimensions. If either $S$ or the normalized size, $\lambda$, is small, reconnection with a single X-line occurs in collisional or in collisionless phases. When both $S$ and $\lambda$ are sufficiently large, three new multiple X-line phases appear with magnetic islands.  The dynamics of new current sheets between these islands are determined either by collisional physics or by collisionless physics (see Sec.III and Sec.IV.). The conditions for electron runaway are shown as red lines (see Sec. V). The locations for reconnection in Earth's magnetosphere, solar corona, solar chromosphere, and solar tachocline are also shown. The existing experiments, such as Magnetic Reconnection Experiment (MRX), do not have accesses to these new phases. A next generation reconnection experiment (NGRX) is required for such accesses to these new phases directly relevant to reconnection in heliophysical and astrophysical plasmas.}
   \label{diagram}
\end{center}
\end{figure}

The collisional MHD description provides a good description of magnetic reconnection for plasmas in which all the resistive layers remain larger than the relevant ion kinetic scale.
%In contrast, much less explicit attention has been received with regard to the effects of plasma size, which can be actually significant.
%$S$ remains unchanged as long as both $L$ and $\eta$ in Eq.(\ref{Lund}) vary by a same factor while 
%the reconnection process can actually be qualitatively different. 
For example, without a guide field ({\it i.e.} anti-parallel reconnection), 
the transition between collisional and collisionless reconnection occurs~\cite{ma96,cassak05,yamada06,simakov08,malyshkin08} 
when the current sheet half thickness predicted by the Sweet-Parker model approaches
\begin{equation}
\delta_{SP}\equiv{L_{CS} \over \sqrt{S}} = d_i
\label{delta}
\end{equation}
where $d_i \equiv c/\omega_{pi}$ is the ion skin depth. By properly varying both $L$ (and thus $L_{CS}$) and $\eta$ 
(through changing {\it e.g.} electron temperature), $S$ and $d_i$ can be kept constant while 
the relative magnitude of $\delta_{SP}$ to $d_i$ can be reversed, leading to dramatic differences in the structure of the reconnection layer along with clear changes in the magnitude and scaling of the reconnection rate.
This qualitative change can be characterized by the effective plasma size which is defined by
\begin{equation}
\lambda \equiv {L \over d_i}
%\label{lambda}
\end{equation}
so that the second equality in Eq.(\ref{delta}) can be written as
\begin{equation}
S = \epsilon^2 \lambda^2.
\label{cc}
\end{equation}
In the case of a finite guide field, the transition occurs~\cite{kleva95,rogers01,cassak07,egedal07} when $\delta_{SP} = \rho_s$ where $\rho_s \equiv \sqrt{(T_i+T_e) m_i}/q_i B_T$ is ion sound gyroradius, $T_e$ and $T_i$ are electron and ion temperatures, $B_T$ is the total magnetic field including both the reconnecting and guide components, and $m_i$ and $q_i$ are the ion mass and charge. 
In the case of anti-parallel reconnection with upstream plasma $\beta_{up} \ll 1$, $\rho_s$ will be equal to $d_i$ by the virtue of the force balance across the current sheet, 
if the reconnecting magnetic field and the temperatures at the current sheet center are used to calculate $\rho_s$. 
(We note that when $\beta_{up} \gg 1$, the transition scale for $\delta_{SP}$ is less clear since $\rho_s (\simeq \sqrt{\beta_{up}} d_i)$ is separated from $d_i$.)
Therefore, the boundary between collisional and collisionless reconnection is defined by Eq.(\ref{cc}) 
regardless of the presence of a guide field when the definition of plasma effective size is modified to
\begin{equation}
\lambda \equiv {L \over \rho_s}.
\label{lambda}
\end{equation}
Thus, the collisional and collisionless reconnection phases are distinguished in the parameter space of $(\lambda,S)$.
%rather than solely by $S$ alone as sometimes assumed. 
This is illustrated as the black line in the phase diagram in Fig.~\ref{diagram} assuming $\epsilon =1/2$. 
%We note that a previous version of this diagram was discussed using electron runaway conditions~\cite{daughton11a}, see Sec.V later for more detailed discussions.
%(It is worth noting that the above prescription has been developed  based on the well established 
We note that the term \lq\lq collisionless reconnection" is used in this paper for the reconnection process dominated
by the effects beyond collisional MHD, such as two-fluid effects, ion and electron kinetic effects.
Among these, the electron kinetic effects should become important in a similar parameter space defined by the black line as shall be discussed in Sec.V.

\section{Single and multiple X-line collisional reconnection}

For plasmas larger than those specified by Eq.(\ref{cc}), it would appear that the collisional MHD 
description might be valid despite the large $S$. It shall become clear later in Sec.III.B.,
however, collisional models are not sufficient for describing reconnection in these regimes.
Discussions on collisional reconnection have been long dominated 
by debates between the Sweet-Parker model~\cite{sweet58,parker57} 
and the Petschek model~\cite{petschek64}, both of which, however, are unsatisfactory. 
The Sweet-Parker model has been verified numerically~\cite{biskamp86,uzdensky00} and experimentally~\cite{ji98} 
at relatively small values of $S$, 
but it predicts reconnection rates too slow to be consistent with observations of larger $S$ plasmas.
On the other hand, the Petschek model, invoking slow-mode shocks, predicts rates consistent with observations but it requires a localized
resistivity enhancement in simulations~\cite{ugai77,hayashi78,scholer89,biskamp01} and has not yet been verified experimentally.
The origin of the localized resistivity enhancement is hypothesized to be kinetic in nature, but the underlying mechanisms still remain illusive.
While signatures of slow-mode shocks have been reported in the Earth's distant magnetotail~\cite{feldman87}, 
large-scale hybrid (kinetic ions and fluid electrons) simulations have revealed significant discrepancies in the expected 
structure of the discontinuities due to the strong ion temperature anisotropy that is naturally generated in these configurations~\cite{kraussVarban95,lottermoser98}.

There is a growing body of work that suggests it may be necessary to move beyond these 
steady-state models in order to understand the dynamics of magnetic reconnection in large-scale collisional plasmas.
In particular, for sufficiently high Lundquist numbers, resistive MHD simulations feature highly elongated 
layers which breakup into muliple X-lines separated by magnetic islands (or plasmoids)~\cite{bulanov79,syrovatskii81,biskamp82,lee85,matthaeus85,biskamp86}.
These multiple-X line models are inherently time dependent, often generating impulsive reconnection
consistent with observations such as Flux Transfer Events (FTE)~\cite{russell79}.
The plasmoid-like structures are also observed in Earth's distant magnetotail during substorms~\cite{hones84} 
and in the current sheet during solar Coronal Mass Ejections (CME)~\cite{sheeley02}.
Although the multiple-X line models were also applied to explain these observed plasmoids in the magnetotail~\cite{birn80,hautz87,otto90} or on the solar surface~\cite{forbes91,kliem00,shibata01}, they did not receive
much attention until recent theory~\cite{loureiro07} and numerical simulations~\cite{lapenta08,barta08,samtaney09,daughton09b,bhattacharjee09,nishida09,shen11} offered detailed predictions concerning the break-up of Sweet-Parker layers to the so-called plasmoid instability, which produces numerous secondary magnetic islands. 
Although the time-averaged rates can be still different~\cite{bhattacharjee09,shepherd10,uzdensky10} 
depending on the detailed divisions within (see Sec.~III below),
all of them are definitely much faster than the Sweet-Parker rate [Eq.(\ref{SP})]. Thus, the multiple X-line reconnection,
associated with the plasmoid instability, constitutes a new reconnection phase within the collisional reconnection regime. 
%Usually, larger the noise amplitudes the lower $S_c$ is since  instabilities have to grow from some finite fluctuations, which always exist in real plasmas.
Recent MHD simulations indicate that the critical Lundquist number, $S_c$, for the onset of the plasmoid instability is approximately
\begin{equation}
S_c \sim 10^4,
\label{Sc}
\end{equation}
shown by the green line in Fig.~\ref{diagram}. 
However, the precise value of $S_c$ probably depends on the level of pre-existing fluctuations in the plasma~\cite{matthaeus85,daughton09b,loureiro09,huang10}.
We note that the green line is necessarily stopped at the
low $\lambda$ end defined by Eq.(\ref{cc}) due to the invalidity of MHD models in the collisionless phase.

\section{Multiple X-line collisional and collisionless reconnection}

Until quite recently, the boundary between collisional and collisionless reconnection was thought to be given by Eq.(\ref{cc}). 
This may not be true anymore
when the current sheet is unstable to the plasmoid instability, forming thinner current sheets which
may be further subject to new plasmoid instability leading to yet thinner current sheets in a hierarchical fashion 
as proposed by Shibata and Tanuma~\cite{shibata01}.
Eventually, these new current sheets can approach the ion kinetic scales
triggering collisionless reconnection as recently demonstrated by full kinetic simulations with a Fokker-Planck treatment of Coulomb collisions~\cite{daughton09b}. Therefore, the multiple X-line phase can be further
divided into a phase involving only collisional physics (i.e. purely resistive MHD) and a phase involving both collisional and collisionless physics, which we denote as the multiple X-line hybrid phase.

\begin{figure}
\begin{center}
\includegraphics[width=4.5in]{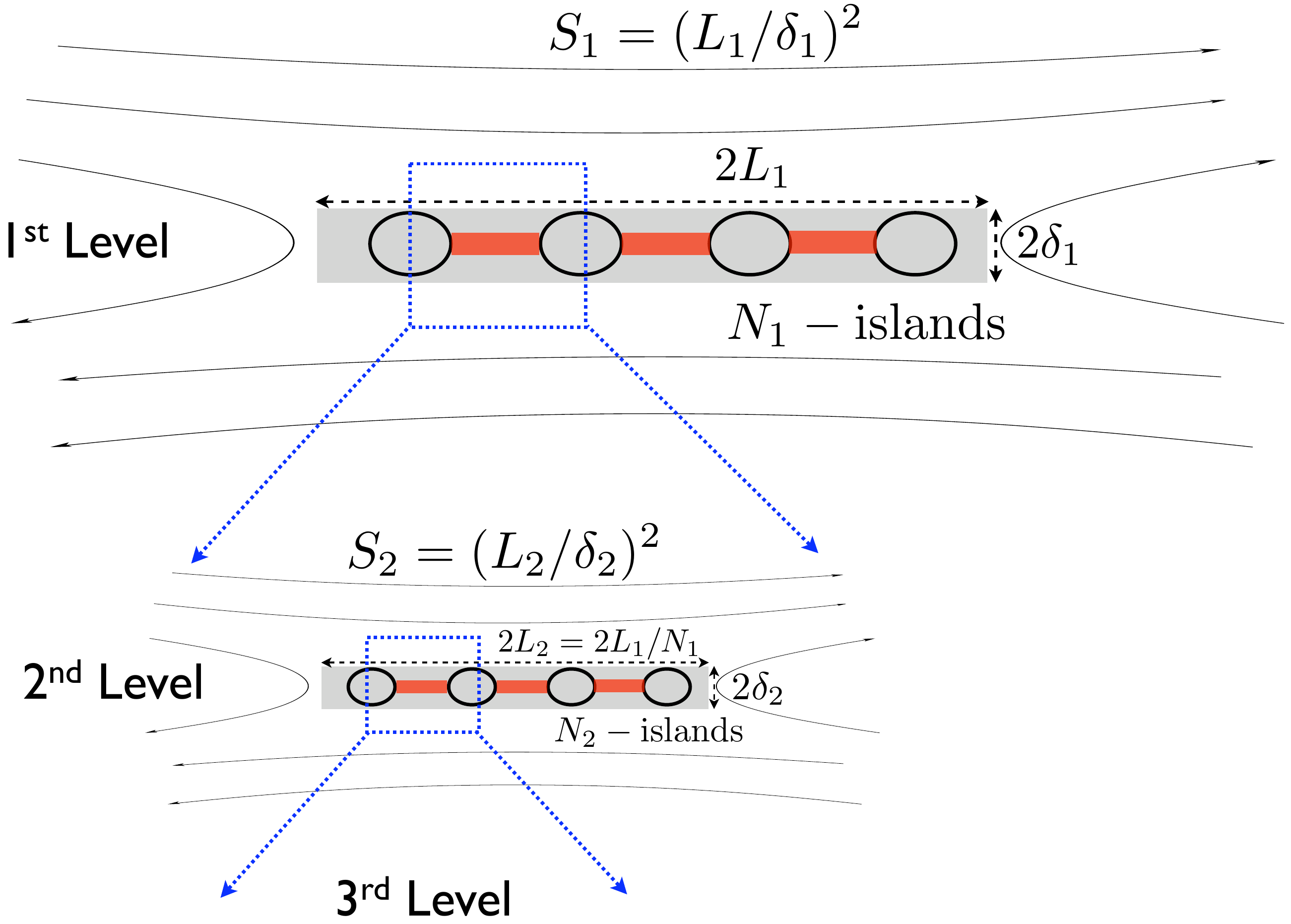}
\caption{In the regimes of high Lundquist number and large plasma size, the plasmoid instability gives rise to a hierarchy of interacting current sheets and islands. The above sketch gives the notation used here for describing this hierarchy.} \label{hierarchy}
\end{center}
\end{figure}

The boundary line in the $(\lambda,S)$ space between these two sub-phases depends on
the detailed physics of unstable current sheets. The main uncertainty originates from the question of
how many islands, on the average, remain within the unstable current sheet at any given time.
% during the nonlinear, highly dynamical process. 
According to linear analytic theory~\cite{loureiro07,ni10}, the number of secondary islands scales as
\begin{equation}
N \sim \left( {S\over S_c} \right)^\alpha,
\label{N}
\end{equation}
where $\alpha=3/8$. This linear prediction has been carefully verified in simulations designed to study the initial breakup of the Sweet-Parker layers \cite{samtaney09,huang10}.  However, nonlinearly many more islands are observed in the simulations\cite{daughton09b,bhattacharjee09,cassak09,huang10} corresponding to scaling parameters in the range $\alpha=0.6 \rightarrow 1$.
A possible explanation for this discrepancy is that, in these nonlinear simulations, the break-up of the original Sweet-Parker layer leads to new current sheets between the islands which are also unstable to the same plasmoid instability as illustrated in Fig.~\ref{hierarchy}, and the islands on more than one level in the hierarchy were counted (see below for more discussions).

At the present time, it appears that there are only two ways to terminate the downward progression in this hierarchy:  (1) either the local Lundquist number of the new current sheets falls below the critical value for the plasmoid instability or (2) the new current sheets approach the ion $\rho_s$ scale where 
collisionless effects dominate.  One can make quantitative predictions regarding these two possible outcomes with just a few simple assumptions.  

\subsection{Multiple X-line collisional reconnection}

As illustrated in Fig.~\ref{hierarchy}, we start by defining the half-length and thickness of the top level Sweet-Parker by $L_1 \equiv L_{CS}$ and 
$\delta_1 \equiv \delta_{SP}$ [Eq.(\ref{delta})], corresponding to a macroscopic Lundquist number of $S_1 \equiv S$ [Eq.(\ref{Lund})]:
$ S_1 = ( L_1/\delta_1)^2$.
At this top level, the development of the plasmoid instability gives rise to $N_1 = (S_1/S_{c})^\alpha$ islands, which breaks the original layer into new sheets with length $L_2=L_1/N_1$.  We assume these new layers are governed by the Sweet-Parker scaling relationships and are also susceptible to the plasmoid instability in the same manner.  Then, the number of islands generated within the second level of the hierarchy is $N_2 = (S_2/S_{c})^\alpha$ where $S_2 = (L_2/\delta_2)^2$ is the Lundquist number of the new sheets, assuming the same reconnection magnetic field strength upstream. Therefore, the $j$th level quantities in the hierarchy are related to the ($j$-1)th level quantities by
\begin{eqnarray}
L_j & = & {L_{j-1}\over N_{j-1}}\\
S_j & = & {S_{j-1}\over N_{j-1}}\\
\delta_j \equiv {L_{j}\over \sqrt{S_{j}}} & = & {L_{j-1}\over \sqrt{S_{j-1}}} {1\over \sqrt{N_{j-1}}} = {\delta_{j-1}\over \sqrt{N_{j-1}}},
\label{Newdelta}
\end{eqnarray}
where the number of islands in the $j$th level, $N_j$, is given by a recursion relation,
\begin{eqnarray}
N_j & = & \left({S_j\over S_c}\right)^\alpha =  \left({S_{j-1}\over S_c}\right)^\alpha \left( S_j \over S_{j-1}\right)^\alpha \nonumber \\
& = & N_{j-1}^{(1-\alpha)}=N_{j-2}^{(1-\alpha)^2}=...=N_1^{{(1-\alpha)}^{j-1}}.
\label{recursion}
\end{eqnarray}
If we terminate the hierarchy at the $j$th level, then the total number of islands in the system ${\cal N}_j$ corresponds to the product of the islands in all the levels
\begin{eqnarray}
{\cal N}_j & \equiv & N_1 N_2 N_3 \; . . .\; N_j \nonumber \\ 
& = & N_1 N_1^{(1-\alpha)} N_1^{(1-\alpha)^2} \; . . .\; N_1^{(1-\alpha)^{j-1}} \nonumber \\
& \equiv & N_1^{\beta_j} = \left( \frac{S_1}{S_c} \right)^{\alpha \beta_j}
\end{eqnarray}
where 
\begin{equation}
\beta_j \equiv  \sum_{n=0}^{j-1} \left( 1 - \alpha \right)^n ={1-(1-\alpha)^j\over \alpha}.
\end{equation}
Note that as the hierarchy becomes deeper $ j \gg 1$ it converges towards $\beta_\infty =\alpha^{-1}$.   
We can now conveniently express the scaling for the total number of islands up through $j$th level in terms of the global Lundquist number
\begin{equation}
{\cal N}_j = \left( \frac{S_1}{S_c} \right)^{1-(1-\alpha)^j}, 
\label{Nscaling}
\end{equation}
while the Lundquist number of the new current sheets at the $j$th level is
\begin{eqnarray}
S_j & = & {S_{j-1}\over N_{j-1}} = {S_{j-2} \over N_{j-1}N_{j-2}} = ... \nonumber \\
&=& {S_1 \over N_{j-1}N_{j-2} ... N_1}=
{S_1 \over {\cal N}_{j-1}} = S_1^{(1-\alpha)^{j-1}} S_c ^{1-(1-\alpha)^{j-1}} \;.
\end{eqnarray}
Notice that for $\alpha <1$ this result implies $S_j > S_c$ for any finite level in the hierarchy, which implies the new levels are always unstable to the plasmoid instability!  Strangely enough, this implies the hierarchy has infinite depth for $\alpha < 1$, and only terminates in the limit $j \rightarrow \infty$ where ${\cal N}_\infty  = S_1/S_c$ and $S_\infty = S_c$.  However, our basic scaling assumption for the number of islands at each level $N_j = (S_j/S_c)^\alpha$ becomes invalid when $N_j$ decreases towards unity. To make a reasonable estimate for the number of levels in the hierarchy, one should consider a cutoff $N_j \sim N_{min}$ below which continuous scaling arguments are meaningless.  
For example, by setting $N_j \ge N_{min}$ in Eq.~(\ref{recursion}) as a cutoff, the maximum level in the hierarchy is
\begin{equation}
j_{max} = 1 + \frac{\ln(\ln(N_{min})) - \ln(\alpha\ln(S_1/S_c))}{\ln(1-\alpha)} \;,
\label{jscaling}
\end{equation}
which diverges logarithmically as $N_{min} \rightarrow 1$.  Picking some reasonable cutoff $N_{min} \sim 2$ and assuming $\alpha=3/8$ will terminate the hierarchy fairly quickly ($j_{max} = 2 \rightarrow 8$) for most any conceivable Lundquist numbers (see Sec. VI). 

Therefore, the scaling of the total number of islands in the hierarchy [Eq.~(\ref{Nscaling})] 
depends on the maximum number of levels, $j_{max}$ [Eq.~(\ref{jscaling})], 
and the island number scaling power index from one level to the next level, $\alpha$.
Applying this estimate to the reported numerical MHD simulations \cite{huang10} yields $j_{max} \sim 3$,
using $\alpha=3/8$. This leads to the predicted scaling of  ${\cal N}_{j_{max}} \sim (S_1/S_c)^{0.76}$. 
However, the linear scaling
of $\alpha=3/8$ does not necessarily apply in the hierarchy model where the nonlinear evolution of islands at one level 
is required to generate new current sheets for the islands at the next level. 
Using $\alpha=0.8$, for example, leads to $j_{max} \sim 2$ and ${\cal N}_{j_{max}} \sim (S_1/S_c)^{0.96}$.
These scalings are not very far from the reported linear scaling of $S^{\sim 1}$ 
given the large uncertainties that still exist (see Fig.~5 in Ref.~51). 
As the hiearchy becomes increasingly deep, the precise value of $\alpha$ no longer matters and the result approaches the linear scaling of $S$ as evident from Eq.(\ref{Nscaling}), consistent with earlier heuristic arguments~\cite{daughton09b,huang10,uzdensky10}.

\subsection{Transition to multiple X-line collisionless reconnection}

\begin{figure}
\begin{center}
\includegraphics[width=4in]{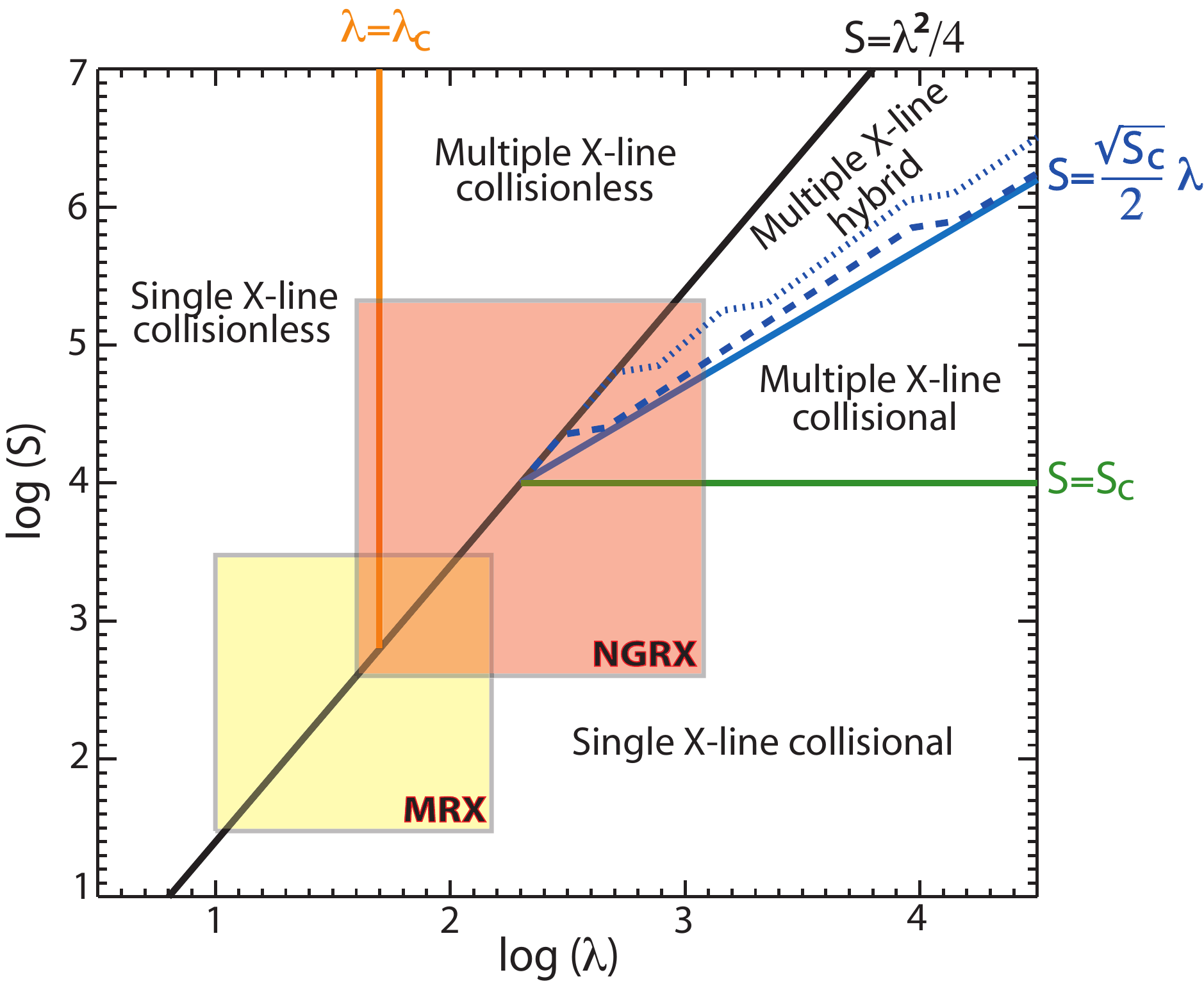}
 \caption{The phase diagram in a smaller parameter space to show dotted and dashed lines better (see texts in Sec. III). Other symbols are same as in Fig.\ref{diagram}.}
 \label{small}
\end{center}
\end{figure}

There is a second way to terminate the downward progression in the hierarchy before reaching the maximum level estimated by Eq.~(\ref{jscaling}).
As discussed in Sec.~I, this occurs when the thickness of a current sheet at a level $j$ ($\geq1$), given by Eq.~(\ref{Newdelta}), 
approaches the ion sound radius to trigger collisionless reconnection:
\begin{eqnarray}
\delta_{j} & = & {\delta_{j-1} \over \sqrt{N_{j-1}}} = {\delta_{j-2} \over \sqrt{N_{j-1}N_{j-2}}} =\;...\;\nonumber\\
& = &  {\delta_1\over \sqrt{N_{j-1}N_{j-2}... N_1}}={\delta_1\over \sqrt{{\cal N}_{j-1}}}=\rho_s.
\label{deltaj}
\end{eqnarray}
Using Eq.~(\ref{Nscaling}), this can be expressed in terms of the critical global Lundquist number, $S=S_1$, as a function of plasma size, $\lambda$, as
\begin{equation}
S=S_c^{(1-c)/2} \left({\epsilon \lambda} \right)^{1+c},
\label{hybrid}
\end{equation}
where
\begin{equation}
c = {(1-\alpha)^{j-1} \over 2-(1-\alpha)^{j-1} } \nonumber\\
\end{equation}
which vanishes in the limits of $\alpha=1$ or $j =\infty$ for $0<\alpha<1$.  
The blue line in Fig.~\ref{diagram} shows the transition boundary in the $(\lambda, S)$ space from multiple X-line collisional phase to multiple X-line hybrid phase in these limiting cases, 
\begin{equation}
S=\sqrt{S_c}\epsilon \lambda
\label{hybrid-two}
\end{equation}
where $S_c=10^4$ and $\epsilon=1/2$, consistent with the previous heuristic arguments~\cite{huang10,huang11}.

For more realistic scenarios with $\alpha <1$ and a finite number of levels, the appxoimate transition boundary can also be estimated.
For any given $S(>S_c)$, the deepest level, $j_{max}$, in the hierarchy is 
given by Eq.~(\ref{jscaling}). Then using this $j_{max}$, the maximum $\lambda$ for 
current sheets at the deepest level in the hiearchy to remain collisional
is determined by Eq.~(\ref{hybrid}).
Examples for $\alpha=3/8$ and $\alpha=0.8$ are shown in Fig.~\ref{small} as a dotted
line and a dashed line, respectively. The steps in these lines correspond to the increases in
$j_{max}$, but overall they are only slightly above the limiting cases of $c=0$.
Lastly, we note that the physics across this boundary is vastly different: on the collisional side
the reconnection is completely determined by collisional MHD physics while on the hybrid
side, both collisional and collsionless physics is important. 
This is consistent with the boundary determined by electron runaway conditions 
which also indicate the change in the required physics (see Sec.V later).
The reconnection rate, on the other hand,
is given by the Sweet-Parker rate of $S_c^{-1/2}\sim 0.01$ on the collisional side while on the hybrid side, the reconnection is faster, but not by a large amount, at the collisionless rate of $0.01-0.1$.

\section{Single and multiple X-line collisionless reconnection}

The last part in our phase diagram concerns the fact that the single X-line current sheet in the collisionless
phase [defined by Eq.(\ref{cc})] may be also subject to secondary collisionless tearing instability.
Unlike the MHD counterpart, we are not aware of any analytic work in this area although there have been
numerical demonstrations~\cite{daughton06,drake06,drake10} and some observational evidence~\cite{chen08}.
In the collisionless limit, sufficiently large kinetic simulations suggest~\cite{daughton06} that the critical size for the secondary island 
formation in the extended current sheet as a result of nonlinear evolution is,
\begin{equation}
\lambda=\lambda_c \sim 50,
\end{equation}
which is shown as the orange line in Fig.~\ref{diagram}.

In principle, multiple X-line collisionless reconnection can also occur in the hybrid phase if the effective plasma
size is larger than $\lambda_c$ at the hierarchy level when the current sheet thickness reaches $\rho_s$. But it
turns out that the condition for such transition is almost identical to Eq.~(\ref{hybrid-two}) as shown below.
The effective plasma size at the $j$th level in the hierarchy, $\lambda_j$, is given by
\begin{equation}
\lambda_j={L_j \over \rho_s}={L_1 \over \rho_s}{1\over {\cal N}_{j-1}}=\epsilon \lambda \left({S_c \over S}\right)^{1-(1-\alpha)^{j-1}}.
\end{equation}
Equating this to $\lambda_c$ yields
\begin{equation}
\epsilon {\lambda \over \lambda_c} = \left({S \over S_c}\right)^{1-(1-\alpha)^{j-1}}
\end{equation}
which reduces to $S=(S_c\epsilon/\lambda_c) \lambda$ in the limiting cases of $\alpha=1$ or $j=\infty$. This condition
is different from Eq.~(\ref{hybrid-two}) by only a factor of 2 when $S_c=10^4$ and $\lambda_c=50$. Thus, the parameter space
for the multiple X-line collisionless reconnection is very limited within the hybrid phase.

\section{Electron runaway condition}

In real plasmas, the whole concept of collisional resistivity (and thus Lundquist number) is restricted to parameter regimes in which the reconnection electric field is small in comparison to the Dreicer runaway limit~\cite{dreicer59} given by
\begin{equation}
E_D \equiv {\sqrt{m_eT_e}\nu_{ei}\over e},
\end{equation}
where $\nu_{ei}$ is electron-ion collision frequency.
%It could be insightful to compare those conditions onto the current phase diagram.
To put the phase diagram into better perspectives with regard to its previous version~\cite{daughton11a}, 
it is important to understand where runaway conditions are unavoidable and how these boundaries correlate with kinetic-scale transitions already discussed.
For a Sweet-Parker reconnecting current sheet, the reconnection electric field is given by
\begin{equation}
E_R=\eta j = {m_e \nu_{ei} \over e^2 n} {B_R \over \mu_0 \delta_{SP}}
\label{E_R}
\end{equation}
where $\eta$ is classical resistivity, $j$ is the peak current density, and $B_R$ is the reconnecting field component.
Using the relation for the plasma $\beta$
\begin{equation}
\beta=2 \left( {\rho_s\over d_i} \right)^2
\end{equation}
where $T_e=T_i$ is assumed, we have
\begin{equation}
{E_R\over E_D} = {2\sqrt{2}\over \beta} {B_R\over B_T} \sqrt{m_e\over m_i} {\rho_s\over \delta_{SP}}.
\end{equation}
For the single X-line reconnection phase, $\delta_{SP}$ is given by Eq.({\ref{delta}), and $E_R/E_D=1$ leads to
\begin{equation}
S={\beta^2 \over 128} \left( {B_T\over B_R} \right)^2 {m_i\over m_e} \lambda^2
\end{equation}
where $B_T$ is the total magnetic field. Assuming $\beta=0.01$, $B_T/B_R=10$, $m_i/m_e=1836$, the above
equation becomes
\begin{equation}
S=0.14 \lambda^2
\end{equation}
which is below the boundary line defined by Eq.(\ref{cc}) but within only a factor of 2 when $\epsilon=1/2$. Therefore, the electron
runaway condition is well coincident with the boundary line between collisional and collisionless phases.

For the multiple X-line reconnection phase, we can use Eq.(\ref{deltaj}) with $\alpha=1$ or $j=\infty$ for simplicity, yielding
\begin{equation}
S={\beta \over 8\sqrt{2}} {B_T\over B_R} \sqrt{m_i\over m_e} \sqrt{S_c} \lambda.
\end{equation}
Using the same example parameters as above, we have
\begin{equation}
S=0.38 \sqrt{S_c} \lambda
\label{runaway}
\end{equation}
which is again within a factor of 2 from Eq.(\ref{hybrid-two}) when $c=0$ and $\epsilon=1/2$. Red lines in Fig.~\ref{diagram} illustrate the
boundaries for electron runaway conditions, which separate collisional and collisionless reconnection phases
for both single X-line and multiple X-line geometries.
The significance of the red lines in the phase diagram is that they separate the regime where reconnection can be described by collisional physics alone from the regime where collisionless physics is required. The alignment of red lines with either the black line or blue line is
consistent with the transitions from collisional reconnection to collisionless reconnection, regardless whether it takes the form of the single X-line or multiple X-lines.

Besides MHD models and fully kinetic models, Hall MHD models and hybrid models (fluid electrons and kinetic ions)
are often used to study the transitions between collisional and collisionless phases~\cite{ma96,birn01,rogers01,cassak05,cassak07,simakov08,malyshkin08,shepherd10,cassak10,huang11}. 
As demonstrated by the comparative studies between different models\cite{birn01}, Hall MHD models
and hybrid models can capture the qualitative boundaries between these two phases. However,
the coincidence of electron runaway conditions with the transition boundaries 
between collisional reconnection and collisionless reconnection raises questions on the suitability
of these fluid models when they are used to study detailed dynamics near the transitions\cite{cassak05,cassak07,shepherd10,cassak10,huang11}. 
The detailed electron kinetic dynamics become important in these regimes but are not yet accurately treated in fluid models.
In particular, the transition into the multiple X-line hybrid phase unavoidably leads to runaway electric fields ($E>E_D$) as illustrated by the red line in Fig.~\ref{diagram}. Fully kinetic simulations including the collision operator have demonstrated~\cite{daughton09a,roytershteyn10} that the mechanism breaking the frozen-in condition changes rapidly across this transition, from ordinary resistivity in the sub-Dreicer collisional limit ($E \ll E_D$) to off-diagonal terms in the electron pressure tensor for the runaway regime.  Once this transition to runway electric fields has occurred, it is unlikely to be reversed as suggested by Hall MHD models\cite{cassak05,huang11} until fast reconnection has depleted the available flux.  Indeed, large-scale collisional kinetic simulations~\cite{daughton09b} have demonstrated that resulting electron layers in this runaway regime can become highly extended and are unstable to secondary magnetic islands in a manner similar to previous collisionless simulations~\cite{daughton06,klimas10}. Further insights emerge when we
divide $S$ by $\lambda$, yielding
\begin{equation}
{S \over \lambda} = \epsilon {\mu_0 \rho_s V_A \over \eta}, % \equiv \epsilon S_{\rho_s},
\end{equation}
which is simply the Lundquist number based on $\rho_s$. 
It has been suggested that there exists a critical value of $S/\lambda \sim 50$ where the dynamics 
can revert from Hall dynamics (kinetic) back into the Sweet-Parker regime~\cite{cassak05,huang11}. 
However, notice that Eq.(\ref{runaway}) 
implies that electron runaway will occur for $S/\lambda > 40$ beyond which simple resistivity models are known to break down.
It remains an outstanding challenge to properly model this dynamics within two-fluid approaches, but comparative studies between these different models should be useful to provide guidance on reliable two-fluid models which can be practically used for the detailed investigations of the phase diagram at large $S$ and $\lambda$ values.

\section{Discussion}

While the simple $S-\lambda$ diagram conveniently summarizes much of the present knowledge regarding the dynamics of magnetic reconnection, there are a variety of other factors that may significantly influence reconnection which have not yet been discussed.  For example, we have largely avoided the onset question ({\it i.e.} how reconnection gets started) which is likely very different for the various parameter regimes.  In particular, the detailed properties of tearing instabilities in various regions of parameter space may play some role in the onset phase of reconnection, but this subject is beyond the focus of the present paper.  As another example, the structure of the large-scale initial condition can also influence the reconnection dynamics, including features such as asymmetric current layers and velocity shear.  There has been some work on both of these issues, but many uncertainties remain. Below we specifically discuss external drive dependence and three-dimensional effects, followed by the discussions on the locations of various plasmas in the phase diagram.

\subsection{Dependence of external drive}

%There are a few caveats which deserve some discussions on the reconnection phase diagram described here.
One might consider that the reconnection process should depend on the external drive or on how much
free energy available in the system for reconnection. This is especially
relevant when reconnection is modeled by a local box around the diffusion region. 
In real applications, the boundary condition may significantly influence the reconnection process within.
%How the boundaries are treated there is often of crucial importance on the reconnection process within. 
However, we point out
that the definition of the Lundquist number given by Eq.(\ref{Lund}) already takes into account of the available
free energy in the system: the half length of the reconnecting current, $L_{CS}$, which is taken as a fraction of
the system size, $L$, is used. If the system is completely relaxed without free energy for reconnection,
then $L_{CS}=0$ even $L$ can be very large. Having $L_{CS}$ on the order of $L$ implies that the available free energy is near its maximum. 
One can imagine a time evolution when the system is driven from its completely relaxed state with $L_{CS}=0$ to $L_{CS} \simeq L/2$ to reach a maximum $S$, 
and the dependence on the drive is actually reflected in the magnitude of $S$ already. 
If the free energy is less than its maximum in a given system, $S$ should be less than its maximum value even $L$ is still same.

\subsection{Influence of realistic three-dimensional dynamics}

The ideas leading to the phase diagram in Fig.~\ref{diagram} are largely based on two-dimensional (2D) models and simulations of reconnection.  At present time, very little is known regarding how reconnection will proceed in large three-dimensional (3D) systems  - either in fluid or kinetic parameter regimes. To begin with, the whole idea of magnetic islands relies upon a high degree of symmetry, which can be achieved in laboratory plasmas (and 2D simulations) but is unlikely to occur in space and astrophysical plasmas.  Instead, extended flux ropes are the natural 3D extension of magnetic islands, and the manner in which these can form and interact is much more complicated than in 2D models.  This is true of both the primary flux ropes which may form due to tearing instabilities, and also secondary flux ropes which can form in the new current sheets that arise from the nonlinear evolution of the primary flux ropes.  The advent of petascale computers over the last few years is permitting 3D kinetic simulations\cite{daughton11b} to explore these ideas for guide field reconnection geometries, where tearing modes can be localized at resonant surfaces across the initial current sheet.  These initial simulations together with Vlasov theory have demonstrated that the spectrum of oblique tearing modes within ion-scale layers is simpler than previously thought\cite{galeev86}, but the resulting flux rope dynamics is still quite rich.  Furthermore, the nonlinear development of the primary flux ropes produces intense electron-scale current sheets near the active x-lines and along the separatrices.   As illustrated in Fig.~\ref{flux-ropes}, in 3D simulations these layers are unstable to the formation of secondary flux ropes over a broad range of oblique angles.  The continual formation and interaction of these flux ropes gives rise to a turbulent evolution that is significantly different than 2D models.  However, the full implication of these results will take years to sort out; researchers are just beginning to scratch the surface.
\begin{figure}
\begin{center}
\includegraphics[width=4.5in]{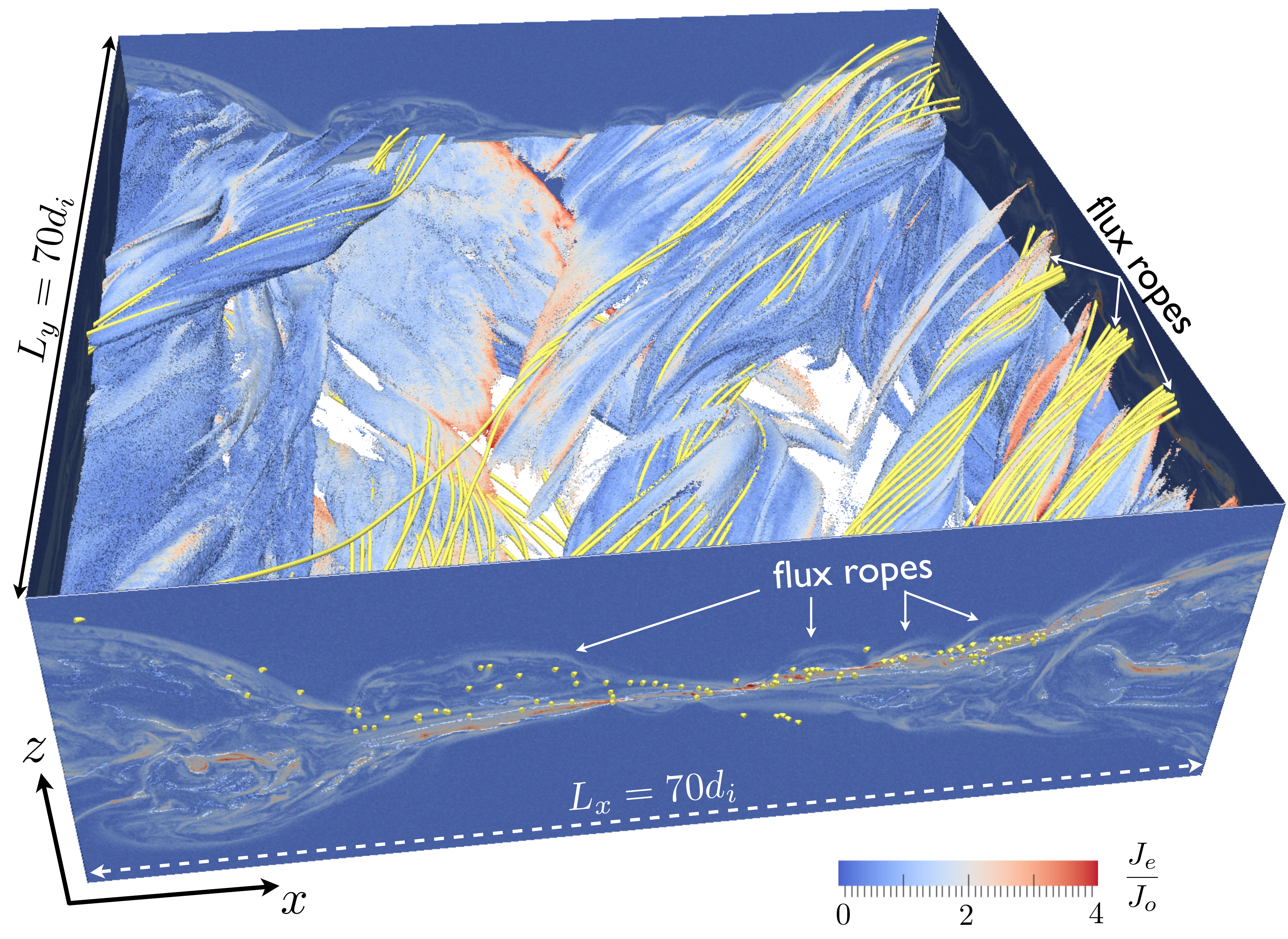}
 \caption{Results from a kinetic simulation\cite{daughton11b} of guide field reconnection showing the formation and interaction of flux ropes as illustrated by an isosurface of the particle density colored by the magnitude of the current density along with sample magnetic field lines (yellow).   Simulation parameters are $m_i/m_e=100$ with the initial guide field equal to the reconnecting field. The domain size is $70 d_i \times 70 d_i \times 35 d_i$ corresponding to $2048 \times 2048 \times 1024$ cells and $\sim10^{12}$ particles.}
\label{flux-ropes}
\end{center}
\end{figure}

In addition to these fundamental issues associated with island (flux rope) formation, there are a wide range of other processes to consider in 3D which may potentially influence the dynamics of magnetic reconnection.  These processes include the lower-hybrid drift instability, driven by the strong diamagnetic currents, streaming instabilities,  modes driven by either electron or ion velocity shear, and a range of kinetic instabilities driven by temperature anisotropy. Even in MHD, influence of a pre-exiting turbulence on reconnection remains an outstanding issue in both 2D~\cite{loureiro09} and 3D~\cite{lazarian99} studies. Huge challenges remain in understanding the role these various process play in reconnection, and how they might change the phase diagram in Fig.~\ref{diagram}.  One of the most long-standing ideas is that instabilities may modify the dissipation physics within electron-scale regions.  However,  there are other possibilities to consider including non-linear couplings between electron and ion-scale features, or the possibility that these instabilities may seed the formation of new flux ropes.  

\subsection{Reconnection in heliophysical, astrophysical and laboratory plasmas}

\begin{figure}
\begin{center}
\includegraphics[width=6.in]{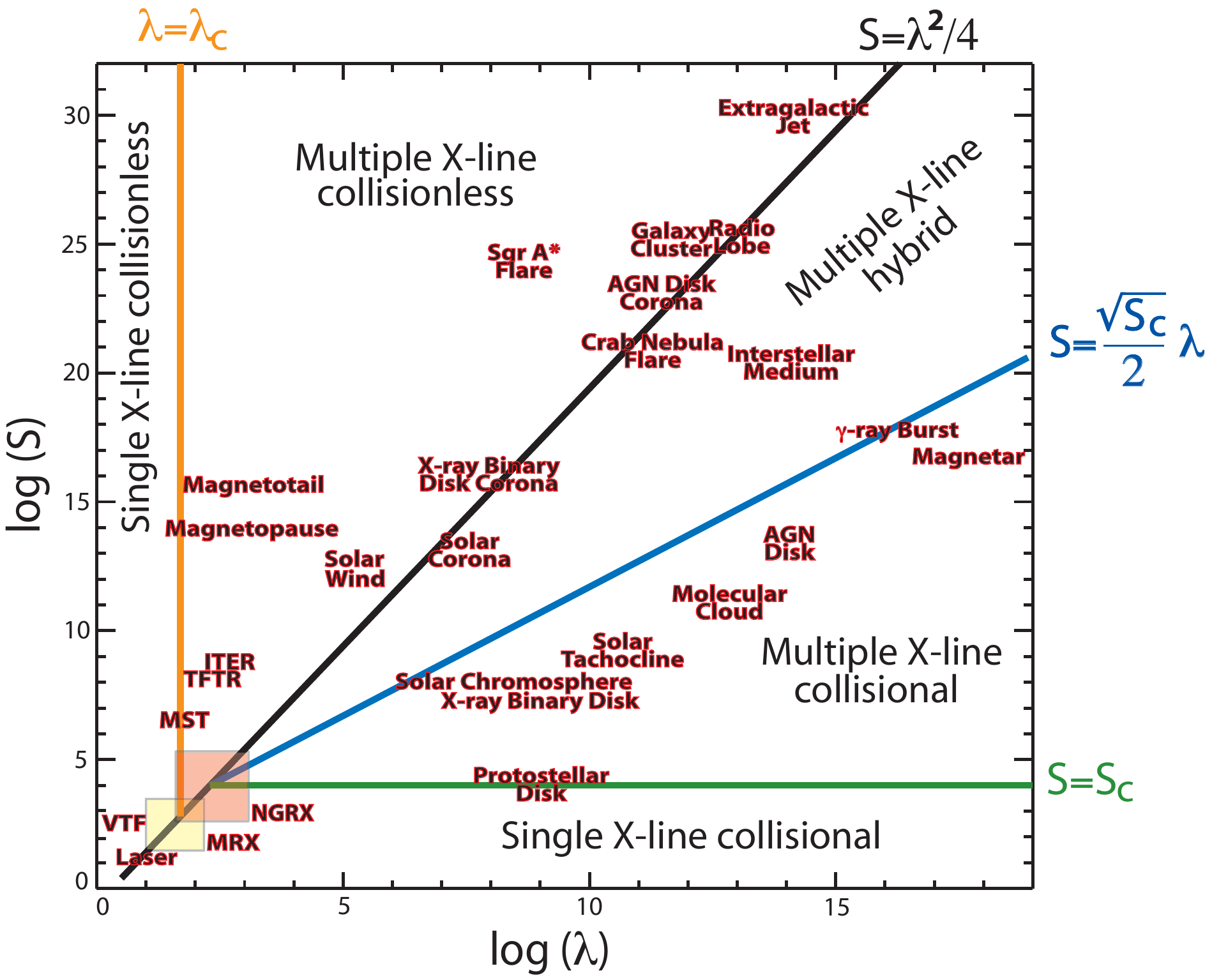}
 \caption{Various laboratory, heliophysical and astrophysical plasmas, in which magnetic reconnection is believed to
 occur, are shown in the phase diagram. Other symbols are same as in Fig.\ref{diagram}. See the text and Table \ref{astro-table} for details.}
\label{large}
\end{center}
\end{figure}

\begin{table}[htdp] %\scriptsize
\caption{Key parameters of various plasmas from laboratory, heliophysics and astrophysics. Unless explicitly stated, assumptions are (1) $\epsilon =1/2$, (2) the reconnecting field is 1/10 of total magnetic field, $B_R=0.1B_T$, (3) equal electron and ion temperatures, $T_e=T_i$, and (4) ions are protons. We note that there are opinions that the plasmas in Crab pulsar wind and radio lobes are nonthermal so that temperature may not be a good description~\cite{uzdensky11b,li11}. There are some laboratory experiments which are not listed: flux rope experiments~\cite{bergerson06,lawrence09,sun10} with $S\sim\lambda\sim10^1$, and plasma merging experiments~\cite{ono96,brown06} with $S=10^2-10^3$ and $\lambda=10^1-10^2$.}
\begin{center}
\resizebox{\textwidth}{!}{
\begin{tabular}{c|p{40mm}|c|c|c|c|c|c|p{60mm}}
location & plasma & size(m) & $T_e$(eV) & $n_e$(m$^{-3}$) & $B_T$(Telsa) & $S$ & $\lambda$ & Notes \\
\hline
& MRX\cite{yamada97b} & 0.8 & 10 & $1 \times 10^{19}$ & 0.1 & $3\times 10^3$ & $1.5\times 10^2$ & $\epsilon=1/4$, $T_i=T_e/2$, $B_R=0.3B_T$ \\
& VTF\cite{egedal07} & 0.4 & 25 & $1.5\times 10^{18}$ & 0.044 & $3 \times 10^2$ & $4\times 10^0$ & $\epsilon=1/4$, $T_i=5$eV, Ar$^+$ \\
& Laser Plasma\cite{nilson08} & $2\times 10^{-4}$ & $10^3$ & $5\times 10^{25}$ & 100 & $2\times 10^1$ & $1\times 10^1$ & Al$^{+13}$, $B_R=B_T$\\
\raisebox{0ex}[0pt]{Lab} & MST\cite{chapman02} & 1.0 & $1.3\times 10^3$ & $9\times 10^{18}$ & 0.5 & $3\times 10^6$ & $6.2\times 10^1$ & $T_i=350$eV, D$^+$, $B_R=0.05 B_T$\\
& TFTR\cite{hawryluk98} & 0.9 & $1.3\times 10^4$ & $1 \times 10^{20}$ & 5.6 & $1 \times 10^8$ & $2.3\times 10^2$ &$T_i=36$keV, D$^+$, $B_R=0.01B_T$ \\ % \#80539
& ITER\cite{johner11} & 4 & $2 \times 10^4$ & $1 \times 10^{20}$ & 5.3 & $6 \times 10^8$ & $5 \times 10^2$ & D$^+$, $B_R=0.01B_T$\\
& NGRX\cite{ji10} & 1.6 & 15 & $3\times10^{19}$ & 0.5 & $2 \times 10^5$ & $1.2\times 10^3$ & $\epsilon=1/4$, $T_i=T_e/2$, $B_R=0.2B_T$ \\
%\cline{2-10}
\hline
& Magnetopause\cite{kivelson95} & $6 \times 10^7$ & 300 & $1 \times 10^7$ & $5 \times 10^{-8}$ & $6 \times 10^{13}$ & $9 \times 10^2$ & $B_R=B_T$, (p.267) \\
& Magnetotail\cite{kivelson95} & $6 \times 10^8$ & 600 & $3 \times 10^5$ & $2 \times 10^{-8}$ & $4 \times 10^{15}$ & $1.3\times 10^3$ & $B_R=B_T$, $T_i=4.2$keV, (p.233) \\
Solar & Solar Wind\cite{kivelson95} & $2 \times 10^{10}$ & 10 & $7 \times 10^6$ & $7 \times 10^{-9}$ & $3 \times 10^{12}$ & $2\times 10^5$ & (p.92) \\
System & Solar Corona\cite{kivelson95} & $1 \times 10^7$ & 200 & $1 \times 10^{15}$ & $2\times 10^{-2}$ & $1 \times 10^{13}$ & $4 \times 10^{7}$ & (p.79)\\
& Solar Chromosphere\cite{malyshkin11} & $1 \times 10^7$ & 0.5 & $1 \times 10^{17}$  & $2\times 10^{-2}$ & $1 \times 10^{8}$ & $3 \times 10^{8}$ & neutral particle effects are weak\cite{malyshkin11}\\
& Solar Tachocline\cite{elliott99,miesch08} & $1 \times 10^7$ & 200 & $1 \times 10^{29}$  & $1$ & $1 \times 10^{9}$ & $5 \times 10^{10}$ \\
\hline
& Protostellar Disks\cite{wardle07} &  $9\times10^9$ & $3 \times 10^{-2}$ & $6 \times 10^8$ & $2\times 10^{-5}$ & $8\times 10^3$ & $1\times 10^9$ & $ L = 2h(R$=1AU), e-n collisions included\cite{malyshkin11}, Mg$^+$\\
& X-ray Binary Disks\cite{frank02,balbus08} & $4 \times 10^4$ & 75 & $1\times 10^{27}$ & 36 & $3 \times 10^7$ & $9 \times 10^8$ & $M=10M_\odot$, $L=2h(R=10^2R_S)$, $\alpha=10^{-2}$, $\dot M=10^{16}g/s$\\
& X-ray Binary Disk Coronae\cite{goodman08} & $3\times 10^4$ & $5 \times 10^5$ & $1\times 10^{24}$ & $1 \times 10^4$ & $1 \times 10^{16}$ & $9\times 10^7$ & $M=10M_\odot$, $R=R_S$, $T_i=(m_p/ m_e)T_e$, $\eta_{Compton}$ included\cite{goodman08}\\
Galaxy & Crab Nebula Flares\cite{tavani11,abdo11,hester08} & $1 \times 10^{14}$ & 130 & $10^6$ & $10^{-7}$ & $5\times 10^{20}$ & $2\times 10^{11}$& pair plasma, $T$ from $B^2_R/2\mu_0=2nT$ \\
& Gamma Ray Bursts\cite{uzdensky11} & $10^4$ & $3\times 10^5$ & $ 2\times 10^{35}$ & $4 \times 10^9$ & $6\times 10^{17}$ & $2 \times 10^{16}$ & pair plasma\\
& Magnetar Flares\cite{palmer05,uzdensky11} &  $10^4$ & $5 \times 10^5$ & $10^{41}$ & $2 \times 10^{11}$ & $6 \times 10^{16}$ & $5 \times 10^{17}$ & pair plasma, SGR 1806-20\\
& Sgr A* Flares\cite{yusef-zadeh09,loeb07} & $2 \times 10^{11}$ & $7\times 10^6$ & $10^{13}$ & $10^{-3}$ & $2 \times 10^{24}$ & $5 \times 10^8$ & $L=2R=20 R_S$ \\
& Molecular Clouds\cite{zweibel99,ferriere01} & $3\times10^{16}$ & $10^{-3}$ & $10^9$ & $2\times 10^{-9}$ & $1\times 10^{11}$ & $7\times 10^{12}$ & neutral particle effects included\cite{malyshkin11}, HCO$^+$ \\
& Interstellar Media\cite{zweibel99,ferriere01} & $5\times 10^{19}$ & 1 & $10^5$ & $5\times10^{-10}$ & $2 \times 10^{20}$ & $1 \times 10^{14}$ & $L$=magnetic field scale height \\
\hline
& AGN Disks\cite{frank02,balbus08,goodman04} & $2 \times 10^{11}$ & 24 & $8 \times 10^{23}$ & 0.5 & $2 \times 10^{13}$ & $1 \times 10^{14}$ & $M=10^8M_\odot$, $L=2h(R=10^2R_S)$, $\alpha=10^{-2}$, $\dot M=10^{26}g/s$\\
\raisebox{-6.0ex}[0pt]{Extra-} & AGN Disk Coronae\cite{goodman08} & $3\times 10^{11}$ & $5 \times 10^5$ & $1\times 10^{17}$ & 4 & $10^{23}$ & $3\times 10^{11}$ & $M=10^8M_\odot$, $R=R_S$, $T_i=(m_p/ m_e)T_e$, $\eta_{Compton}$ included\cite{goodman08}\\
galactic & Radio Lobes\cite{li11} & $3\times 10^{19}$ & 100 & 1 & $5\times 10^{-10}$ & $2 \times 10^{25} $ & $8 \times 10^{12} $\\
& Extragalactic Jets\cite{lapenta05} & $3 \times 10^{19}$ & $10^4$ & $3 \times 10^1$ & $10^{-7}$ & $6 \times 10^{29}$ & $1 \times 10^{14}$ & 3C 303\\
& Galaxy Clusters\cite{kunz11} & $6\times 10^{18}$ & $5\times 10^3$ & $4\times 10^4$ & $ 2 \times10^{-9}$ & $2 \times 10^{25}$ & $6 \times 10^{11}$ & A1835 \\
\end{tabular}}
\end{center}
\label{astro-table}
\end{table}%

Despite these rather serious caveats discussed in the previous section, it is interesting to place plasmas from laboratory, heliophysics and astrophysics in the phase diagram. In Fig.~\ref{diagram}, some heliophysical and laboratory example plasmas are shown. In this section, results from a more extensive survey of astrophysical plasmas are summarized in Table \ref{astro-table} and Fig.~\ref{large} with references from which typical plasma parameters were taken. In general, these parameters are associated with large uncertainties due to limited measurements available from these distant plasmas and crude models used for the estimation. Extreme astrophysical conditions~\cite{uzdensky11}, such as special relativity and radiation, are not taken into account here since these effects on collisionless plasmoid instabilities is just beginning to be explored numerically~\cite{jaroschek09,liu11}. On the log-scales as in Fig.~\ref{large}, however, even an order of magnitude of the uncertainty does not change the location of these plasmas by much in the phase diagram.

The cases shown in the phase diagram can be roughly grouped into three groups. The first group includes high temperature fusion plasmas and Earth's magnetospheric plasmas. These plasmas are completely in the collisionless phases, either with a single X-line or multiple X-lines, depending on whether the plasma effective sizes, $\lambda$, are larger than the critical $\lambda_c$. The plasma for the Sgr A* flares may also belong to this group.

The second group of plasmas cluster along the black line separating multiple X-line collisionless phases and multiple X-line hybrid phase. It spans over huge ranges from solar corona, accretion disk coronae, Crab nebula flare, to galaxy clusters, radio lobes, and extragalactic jets. When $S$ and $\lambda$ are both small, this same line separates single X-line collisional phase and collisionless phase. With a single X line, the reconnection in the collisional phase was known to be much slower than in the collisionless phase. The plasma collisionality was argued~\cite{uzdensky07b} to regulate itself so that the plasma always stay near the marginal collisionality, based on the reasoning that fast reconnection should effectively release magnetic energy evaporating nearby dense neutral gases (such as in the solar chromosphere) to increase density and collisionality until reconnection slows to a collisional rate. An alternative model was also proposed~\cite{cassak08} based on self-regulation of electron temperature to maintain marginal collisionality through a similar but different reasoning: higher temperature lowers collisionality and fastens reconnection, and thus depletes quickly available magnetic energy and eventually slows reconnection and cools the plasma while lower temperature increases collisionality and slows reconnection, and thus accumulate magnetic energy and eventually trigger faster reconnection and heat the plasma. 

However, at large $S$ and $\lambda$ values for all plasmas in the second group, the marginality black line now separates multiple X-line collisionless phases and multiple X-line hybrid phase in the phase diagram. Now there is numerical evidence~\cite{daughton09b,shepherd10} that the reconnection rates in multiple X-line hybrid phase are as fast as the single X-line collisionless rate, consistent with the theoretical argument~\cite{uzdensky10} that the global reconnection rate is determined by a dominant reconnection site in the island hierarchy which should be collisionless in the hybrid phase.  Therefore, the self regulation arguments for the collisionality mentioned above do not seem to hold at the large $S$ and $\lambda$ phases, but much work still remains to be done. The accumulation of energetic particle populations is suggested~\cite{drake10} as another player in the self regulation process of reconnection rate in the multiple X-line collisionless phase. Energetic particle populations should be regulated also by finite collisions in the hybrid phase, but detailed dynamical processes need to be investigated.

The third and final group of plasmas shown in Fig.~\ref{large} occupy much of the multiple X-line collisional phase:
accretion disk interiors, solar chromosphere and tachocline, molecular clouds, gamma ray bursts and magnetar flares. It could be argued that they form a line slightly below but along the boundary (blue line) between the multiple X-line collisional and hybrid phases.
It is conceivable that the self-regulation arguments for collisionality~\cite{uzdensky07b,cassak08} could be applied here since collisional reconnection dominates at the deepest level of the hierarchy on the one side of the boundary while collisionless reconnection dominates on the other side. 
In fact,  it has been suggested through Hall MHD simulations~\cite{shepherd10} that reconnection in the multiple X-line collisional phase is much slower than that in the hybrid phase although it is much faster than the single X-line collisional (Sweet-Parker) rate. However, the reconnection rate is not so different: collisionless rates are around 0.1 and while the collisional rates are around $1/\sqrt{S_c} \sim 0.01$ at the deepest level of the hierarchy. A key question here is what determines the overall reconnection rate in a hierarchy of islands 
and whether it is indeed dominated by the reconnection process at the deepest level~\cite{uzdensky10} or by the reconnection process at all levels in an integrated way.

There are two special cases which do not belong to either of the above three groups: protostellar disks and interstellar media.
Protostellar disks have lowest $S$ among the objects we surveyed and are located between the single X-line and multiple X-line collisional phase. Interstellar media are right in the middle of the hybrid phase, and probably both collisional MHD physics and collisionless physics are important in charactering reconnection processes there as a part of the galactic dynamo.

Lastly, we note that currently there are no laboratory experiments which can be used to study all of these new phases of magnetic reconnection. Laboratory experiments have been playing important roles in the reconnection research: confirming some leading theoretical or numerical models such as Sweet-Parker~\cite{ji98} and collisionless reconnection models~\cite{ren05} while challenging others such as Petschek model; benchmarking state-of-the-art numerical simulations~\cite{ji08,dorfman08,roytershteyn10}; discovering 3D phenomena~\cite{carter02a,ji04,fox08,katz10}; studying flux rope dynamics~\cite{lawrence09,sun10}, to name a few. As mentioned above, the main research tool on physics of new reconnection phases is numerical simulations using either full particle, Hall MHD, or resistivity MHD codes. Existing experiments, such as MRX, do not have accesses to these new phases which are important for the emerging themes of particle acceleration by magnetic reconnection~\cite{drake06,drake10}. While numerical simulations, coupled closely with analytic theory, will continue to be a major player at this front, a next generation reconnection experiment (NGRX) based on the MRX concept is considered as a candidate for such a laboratory experiment~\cite{ji10}. The parameter ranges for both MRX and NGRX are also indicated in the phase diagram for their coverages.

\section*{Acknowledgements}

HJ acknowledges support from the U.S. Department of Energy's Office of Science - Fusion Energy Sciences Program, and the Princeton Plasma Physics Laboratory's Laboratory Directed Research and Development Program. WD acknowledges support from the U.S. Department of Energy through the Los Alamos National Laboratory's Laboratory Directed Research and Development Program.  Simulation in Fig. 4 was performed on Kraken with an allocation of advanced computing resources provided by the National Science Foundation at the National Institute for Computational Sciences.
HJ appreciates suggestions on parameters of various astrophysical plasmas and their references by Jeremy Goodman, Hui Li, Alex Schekochihin, Farhad Zadeh, and Ellen Zweibel. We greatly appreciate critical feedbacks from Amitava Bhattacharjee, Ellen Zweibel, an anonymous referee, and especially from Dmitri Uzdensky who read our manuscript carefully and provided a long list of constructive comments.
We are also grateful to Masaaki Yamada and Stewart Prager for valuable discussions. 

\bibliography{reconnection}

%\begin{thebibliography}{10}
%\end{thebibliography}

\end{document}